\documentstyle [aasms4,psfig,epsf,12pt]{article} 

\slugcomment{Send preprint requests to ''raw@cosmos.la.asu.edu'' } 

\lefthead{Windhorst et al.} 
\righthead{Deep $HST/PC$ images of a young radio galaxy at $z=2.390$}

\begin{document} 


%
\def\i {\indent}  
\def\n {\noindent}
\def\no{\noindent}
\def\b {\bigskip} 
\def\m {\medskip}
\def\s {\smallskip} 
\def\bi{\bigskip\indent}     
\def\mi{\medskip\indent}
\def\si{\smallskip\indent}   
\def\bn{\bigskip\noindent}     
\def\mn{\medskip\noindent}
\def\sn{\smallskip\noindent}
\def\cl{\centerline} 
\def\ve{\vfill\eject}
\def\\{$\backslash$}
%
\def\hh       {{$^{h}$}}
\def\mm       {{$^{m}$}}
\def\ss       {{$^{s}$}}                  
\def\deg      {{\ifmmode^\circ\else$^\circ$\fi} } 
\def\arcm     {{\ifmmode {'\ }\else$'     $\fi} } 
\def\arcs     {{\ifmmode{''\ }\else$''    $\fi} } 
\def\sspt     {{$\buildrel{s}           \over .$}}
\def\degpt    {{$\buildrel{\circ}       \over .$}}
\def\arcmpt   {{$\buildrel{\prime}      \over .$}}
\def\arcspt   {{$\buildrel{\prime\prime}\over .$}}
\def\magpt    {{$\buildrel{m}           \over .$}}
\def\A        {{$\rm\AA$} }
\def\ABnu     {{$AB_{\nu}$} }
\def\Al       {{$A_{\lambda}$} } 
\def\AV       {{$A_{V}$} }
\def\aLyr     {{$\alpha$\ Lyr} }
\def\amed     {{$\alpha_{med}$} }
\def\bII      {{$b^{II}$} }
\def\Bj       {{$B_{J}$} }
\def\U        {{$U$} }
\def\B        {{$B$} }
\def\V        {{$V$} }
\def\R        {{$R$} }
\def\I        {{$I$} }
\def\Lyaband  {{Ly$\alpha_{410}$} }
\def\Uband    {{$U_{300}$} }
\def\Bband    {{$B_{450}$} }
\def\Vband    {{$V_{606}$} }
\def\Iband    {{$I_{814}$} }
\def\Rband    {{$\cal R$} }
\def\chisq    {{$\chi^{2}$} }
\def\cf       {{\it cf.} }
\def\cge      {{$_ >\atop{^\sim}$}}
\def\cle      {{$_ <\atop{^\sim}$}}
\def\degsq    {{\ $deg^{2}$} }
\def\sqdeg    {{\ $deg^{-2}$} }
\def\eg       {{\it e.g.}, }
\def\emin     {{\ $e^{-}$} }
\def\eminsq   {{\ $(e^{-})^{2}$} }
\def\EBminV   {{$E_{B-V}$} }
\def\et       {{et\thinspace al.} }     
\def\etal     {{et\thinspace al.} }     
\def\etc      {{\it etc.,\/~}}
\def\ergcms   {{\ $ergs\ cm^{-2}\ s^{-1}$} }
\def\ergs     {{\ $ergs\ s^{-1}$} }
\def\Fl       {{$F_{\lambda}$} }
\def\Fnu      {{$F_{\nu}$} }
\def\g        {{$g$} }
\def\r        {{$r$} }
\def\i        {{$i$} }
\def\Ho       {{$H_{0}$} }
\def\kms      {{\ $km\ s^{-1}$} }
\def\kmsMpc   {{\ $km\ s^{-1}\ Mpc^{-1}$} }
\def\lII      {{$l^{II}$} }
\def\Ha       {{$H\alpha$} }
\def\Hb       {{$H\beta$} }
\def\Hc       {{$H\gamma$} }
\def\Hd       {{$H\delta$} }
\def\hmin     {{$h^{-1}$} }
\def\ie       {{\it i.e.}, }
\def\Jp       {{$J^{+}$} }
\def\leff     {{$\lambda_{eff}$} }
\def\lya      {{Ly$\alpha$} }
\def\Lya      {{Ly$\alpha$} }
\def\Lrad     {{$L_{rad}$} }
\def\LIR      {{$L_{IR}$} }
\def\Lopt     {{$L_{opt}$} }
\def\LUV      {{$L_{UV}$} }
\def\LX       {{$L_{X}$} }
\def\Lstar    {{$L^{*}$} }
\def\Lo       {{$L_{\odot}$} }
\def\Lsun     {{$L_{\odot}$} }
\def\magarc   {{\ mag\ arcsec$^{-2}$} }
\def\MB       {{$M_{B}$} }
\def\MV       {{$M_{V}$} }
\def\MR       {{$M_{R}$} }
\def\MI       {{$M_{I}$} }
\def\Mgas     {{$M_{gas}$} }
\def\Mo       {{\ $M_{\odot}$} }
\def\Msun     {{\ $M_{\odot}$} }
\def\Mstar    {{$M^{*}$} }
\def\muK      {{\ $\mu$K} }
\def\muJy     {{\ $\mu$Jy} }
\def\NH       {{$N_{H}$} }
\def\Nm       {{$N(m)$} }
\def\nref     {\noindent\parshape 2 0.0 truein 06.5 truein 0.4 truein 06.1 truein}
\def\persec   {{\ sec$^{-1}$} }
\def\perster  {{\ sr$^{-1}$} }
\def\Pstar    {{$P^{*}$} }
\def\Pivstar  {{$P_{1.4}^{*}$} }
\def\Piv      {{$P_{1.4}$} }
\def\Pev      {{$P_{8.4}$} }
\def\pp       {\noindent\parshape 2 0.0 truein 06.5 truein 0.4 truein 06.1 truein}
\def\ppi      {\noindent\parshape 1 0.315 truein 06.185 truein}
\def\qo       {{$q_{0}$} }
\def\re       {{$r_e$} }
\def\rs       {{$r_s$} }
\def\rhl      {{$r_{hl}$} }
\def\Siv      {{$S_{1.4}$} }
\def\Sev      {{$S_{8.44}$} }
\def\Snu      {{$S_{\nu}$} }
\def\Sp       {{$S_{p}$} }
\def\Sint     {{$S_{int}$} }
\def\Smap     {{$S_{map}$} }
\def\ssky     {{$\sigma_{sky}$} }
\def\Thmed    {{$\Theta_{med}$} }
\def\Vmax     {{$V_{max}$} }
\def\Wm       {{\ $W\ m^{-2}$} }
\def\WHz      {{\ $W\  Hz^{-1}$} }
\def\WmHz     {{\ $W\ m^{-2}\  Hz^{-1}$} }
\def\WHzm     {{\ $W\  Hz^{-1}\ m^{-2}$} }
\def\Wl       {{$W_{\lambda}$} }
\def\EW       {{$W_{\lambda}$} }
\def\wth      {{$w(\theta$)} }
\def\wtheta   {{$w(\theta$)} }
\def\zest     {{$z_{est}$} }
\def\zmed     {{$z_{med}$} }
\def\zf       {{$z_{f}$} }
\def\zform    {{$z_{form}$} }
%
\def\AAP     {A\&A, }
\def\AAS     {A\&AS, }
\def\AJ      {AJ, }
\def\APJ     {ApJ, }
\def\APJL    {ApJL, } 
\def\APL     {ApJ, }  
\def\APJS    {ApJS, }
\def\APSS    {Ap\&SS, }
\def\ARAA    {ARA\&A, }
\def\AUSJP   {Australian J. Phys., }
\def\BAAS    {BAAS, }
\def\HOA     {Highlights Astr., }
\def\IAU     {IAU Symp., }
\def\MEMRAS  {MmRAS, }
\def\MNRAS   {MNRAS, }
\def\NAT     {Nature, }
\def\PASJ    {PASJ, }
\def\PASP    {PASP, }
\def\PHD     {Ph.D. thesis, }
\def\QJRAS   {QJRAS, }
\def\SCAM    {Sc. Am., }
\def\SCI     {Science, }
\def\SOVAST  {Sov. Astr., }
\def\SPIE    {SPIE, }
\def\VIS     {Vistas in Astronomy, }
\def\VLA     {VLA Test Memorandum, No.}
\def\IAUABD  {in IAU Symposium 124, 
             Observational Cosmology, 
             ed. A. Hewitt, G. Burbidge, \& L. Z. Fang (Dordrecht: Reidel), }

\title{Deep Hubble Space Telescope\altaffilmark{1}/Planetary Camera imaging \\
of a young compact radio galaxy at $z=2.390$ }

\author{Rogier A. Windhorst } 
\affil{Department of Physics and Astronomy, Arizona State University, \\ 
Box 871504, Tempe, AZ\ \ 85287-1504 } 

\author{William C. Keel }
\affil{Department of Physics and Astronomy, University of Alabama, \\ 
Box 870324, Tuscaloosa, AL\ \ 35487-0324 }

\and 

\author{Sam M. Pascarelle }
\affil{State University of New York at Stony Brook \\ 
Astronomy Program, Stony Brook, NY\ \ 11794-2100 }

\altaffiltext{1}{Based on observations with the NASA/ESA {\it Hubble Space
Telescope} obtained at the Space Telescope Science Institute, which is operated
by AURA, Inc., under NASA Contract NAS 5-26555.}

\begin{abstract} 

We present deep 63-orbit $HST/PC$ images at $\sim$0\arcspt 06 FWHM resolution in
the filters \Bband, \Vband \& \Iband --- as well as in redshifted \Lya --- of
the radio source LBDS 53W002, a compact narrow-line galaxy at z=2.390. These
images allow us to distinguish several morphological components: (1) an
unresolved nuclear point source (\cle 500 pc at z=2.390 for \Ho=75, \qo=0),
likely the central AGN which contains \cle 20--25\% of the total light in $BVI$;
(2) a compact continuum core (\re$\simeq$0\arcspt 05); (3) a more extended
envelope with an $r^{1/4}$-like light-profile and \re$\simeq$0\arcspt 25
($\sim$2 kpc); (4) two blue ''clouds'' roughly colinear across the nucleus
aligned with the radio source axis and contained well within the size of the
radio source. $(B-I)$ color maps may suggest a narrow dust lane crossing between
the nucleus and the smaller blue cloud. The radio source is {\it not} smaller
than the distance between the blue continuum clouds, and coincides with a bright
\Lya ''arc'' in the western cloud, suggesting that jet-induced star-formation
could cause both blue clouds, except the outer parts of the western cloud. The
shape of this larger blue cloud suggests reflected AGN continuum-light shining
through a cone (plus re-radiated \Lya in emission). The OVRO interferometric
CO-detection (Scoville \etal 1997) on {\it both} sides of 53W002 --- and in the
same direction as the continuum clouds {\it and} the radio jet --- also suggest
a star-bursting region induced by its radio jet, at least in the inner parts.
Hence, both mechanisms likely play a role in the ``alignment effect''. Even at
radio powers $\sim$1.5 dex fainter than the 3CR sources, we thus find many of
the same aligned features and complex morphology, although at much smaller
angular scales and lower optical--$UV$ luminosities. We discuss the consequences
for 53W002's formation in the context of the 16 sub-galactic objects at
z$\simeq$2.40 around 53W002 (Pascarelle \etal 1996). 

\end{abstract} 



\section{Introduction} 

To understand galaxy formation, one needs to observe high redshift galaxies at
the highest possible resolution and sensitivity. This can be done with the
0\arcspt 0455 pixels of the Planetary Camera ($PC$) in the refurbished $HST$
Wide Field Planetary Camera 2 ($WFPC2$), although sufficient surface brightness
(SB) sensitivity can only be obtained with the $PC$ for relatively compact
objects. Jet-induced star formation or non-thermal radiation scattered in a
reflection cone are the most probable radiation processes in ultraluminous high
redshift 3CR and 1 Jy radio galaxies (Chambers, Miley \& van Breugel 1990;
McCarthy \etal 1991). It is not clear that these processes are universal, and
their role needs to be clarified at high resolution for $\sim$30-100$\times$
weaker radio galaxies. To address these issues, we obtained deep multicolor
high-resolution $HST/PC$ images of the faint (V$\simeq$23.0 mag), compact (\cle
1\arcs), steep spectrum ($\alpha\simeq$1.2), radio (\Siv$\simeq$50 mJy) galaxy
LBDS 53W002 (Windhorst, van Heerde, \& Katgert 1984a) which has {\it narrow}
emission lines at $z=2.390$ (Windhorst \etal 1991; hereafter W91). 53W002's
radio power (log\Piv=27.5 \WHz; using \Ho=75\kmsMpc and \qo=0.0 throughout) is
$\sim$2.5 dex above the FR-I/II break luminosity \Pstar at z=0. However, its
compact radio morphology is {\it not} of FR-II type. Ground-based and
pre-refurbished $HST$ {\it continuum} images (Windhorst, Mathis, \& Keel 1992,
W92) showed some alignment with the radio source axis on 0\arcspt 5--1\arcspt 0
scales (4.5--9 kpc), which itself is aligned with the much larger ground-based
\Lya cloud ($\sim$25$\times$45 kpc; W91). 


Our redshifted \Lya images in the $WFPC2$ medium-band filter F410M showed 16
possible compact \Lya emitters surrounding 53W002 at $z=2.40$ (Pascarelle \etal
1996; P96). Nine of these objects have been spectroscopically confirmed thus
far, possibly a group or cluster in formation (P96; Armus \etal 1997 in prep.;
Keel \etal 1997, in prep.). The $PC$ images were obtained to constrain the
relative contributions from 53W002's AGN and its young stellar population, and
to examine the relations between these components and its dynamics during the
galaxy collapse --- whether this occurred as a global halo collapse (\cf Eggen,
Lynden-Bell, \& Sandage 1962; ELS62), through the rapid merging of many
sub-galactic sized objects (\eg Searle \& Zinn 1978, SZ78; P96), through
jet-induced star-formation (\eg Chambers \etal 1990), or a combination thereof.

\section{$WFPC2$ observations and processing } 

Deep $HST$ exposures were taken with $WFPC2$ in Cycle 4: 12$\times$1700 sec in
the F606W (``$V$'', \leff$\simeq$5940\AA\ ) and F814W (``$I$'', \leff$\simeq$
7920\AA\ ) filters. Two sets of 6$\times$1700 sec each were obtained at two
different locations, taken a few days apart and separated by SAA passages
(Driver \etal 1995a, D95a). In Cycle 5, four sets of 6$\times$2400 sec exposures
were taken in F450W (``$B$'', \leff$\simeq$4520\AA\ ; Odewahn \etal 1996; O96)
and three sets of 5$\times$2700 sec exposures in the medium-band filter F410M
(\leff$\simeq$4090\AA\ or \Lya at z$\simeq$2.36$\pm$0.06; P96). The 48-orbit
stack of $BVI$ images is shown as a color Plate in Fig. 1. The boxes show
enlargements to the same angular scale for 53W002 and for the $z=2.40$ Objects
18 \& 19 of P96 (seen in camera WF2). All calibrations and reductions followed
Driver \et al. (1995b), D95a, O96 \& Windhorst \etal (1994a). A recent 10-hr
8.44 GHz VLA image with the C-array yielded three radio sources associated with
the three AGN in Plate 1 with significant radio-optical offsets
[$\simeq$(+0\arcspt 56,+1\arcspt 92)$\pm$ 0\arcspt 14)]. This allowed us to
bring the high resolution VLA 8.44 GHz (W91) and 15 GHz (Scoville \etal 1997;
S97) images onto 53W002's optical center --- and likely its AGN (Fig. 2, \S 3.1)
--- to within \cle 3 $PC$ pixels. 


\section{Structure of the $PC$ images of 53W002 } 

Here we discuss the various discrete components of this young galaxy as
revealed in the $PC$ images. We emphase that given the fitting errors of the
respective components, and given the limit of the $HST/PC$ resolution, the nature
and quoted fluxes of each component (as fractions of the total flux) are not
necessarily unique or uncorrelated, although the listed components are
apparently necessary to make a full description of 53W002's PC-morphology (see
Fig. 2a--2f, Plate 2):

\subsection {The maximum AGN contribution } 

To constrain the maximum possible AGN contribution to 53W002's continuum, we
subtracted a scaled PSF from the image core. As PSF, we used the faint red star
''S'' noted by W92 (see Fig. 1 here) --- circularly averaged to improve its S/N.
The maximum possible point source that can be subtracted from 53W002's core
without making its central flux negative is 25$\pm$2\% of the total light in
$B$, 21$\pm$3\% in $V$, and 20$\pm$2\% in $I$. This light is contained within
0\arcspt 06 FWHM, or $\sim$500 pc, and has blue colors ($B-I\simeq$0.06 mag), as
expected for an AGN at $z\simeq2.4$. The upper error boundaries of these
fractions are firm, so that the AGN contribution to 53W002's total continuum is
definitely $\le$30\%, but the lower boundaries are soft. The AGN contribution
could be \cle 15\% of its total continuum, which would in Fig. 3 produce
light-profiles that are straighter in $r^{1/4}$-space. These fractions are
consistent with the spectroscopic limits (from C-IV/\Lya and N-V/\Lya ratios) to
the AGN's restframe UV-continuum fraction that is needed to power the
high-ionization part of the Seyfert-like emission lines (\cle 35$\pm$15\%; W91),
and with the 30$\pm$10\% value from the deconvolved Cycle 1 images (W92).
53W002's AGN contribution is: $V^{AGN}\simeq$24.3, or $M_V^{AGN}\simeq$--21.8,
assuming a power-law $K$-correction with $\alpha\simeq$1.0 (W91, P96). 


\subsection {The inner resolved continuum ''core'' } 

To address the symmetric extended component, we only consider and fit the
``clean'' quadrant between the blue clouds discussed in \S 3.4 (Fig. 2a--2f).
The central parts of the galaxy in F450W cannot be simultaneously fit by a
single $r^{1/4}$ or exponential law. Comparing these profile-fits to the PSF
shows that a small additional central light distribution (with \re$\sim$0\arcspt
05) is required, containing about half the flux of the nuclear point source, but
with redder colors ($B-I$=0.8$\pm$0.1).

\subsection {The remaining $r^{1/4}$-like profile } 

After subtraction of the central unresolved AGN component (\S 3.1), the
underlying galaxy can be measured only in the quadrant {\it between} the two
blue clouds (\S 3.4). We concentrate on a region south of the nucleus, where
nearly a full 90$^\circ$ quadrant is clear of these contaminating sources.
Details of the elliptical profile fitting technique and its errors are given by
W92, Keel \& Windhorst (1993), Windhorst \etal (1994b, W94b), Mutz \etal (1997;
M97), and Schmidtke \etal (1997; Sc97). The major source of error is the
sky-subtraction, but the sky in the PC-image stack is sufficiently low and flat
that sky-subtraction errors are \cle few \% of sky. Given these caveats, the
$BVI$ light-profiles (Fig. 3) follow an $r^{1/4}$-like profile closer than an
exponential disk, although an early-type galaxy with a bulge-to-disk ratio \cge
3--5 cannot be ruled out. Most of the deviation from an $r^{1/4}$-law at
$r^{1/4}$\cge 0.75 ($r$\cge 0\arcspt 32) occurs in $B$ \& $V$, and is due to the
faint blue cloud leaking into the uncontaminated quadrant, and not only due to
sky-subtraction errors, which affect the profile for SB$_{BVI}$\cge
25.5--26.0\magarc. We fit a family of $r^{1/4}$-profiles --- convolved with the
PSF (\S 3.1) --- to the $PC$ data in this area over the radial range $r$=2--9
pixels (0\arcspt 1--0\arcspt 4). The best $r^{1/4}$-fit has (a/b)=1.25$\pm$0.1
and \re$\simeq$0\arcspt 20$\pm$07 in $B$ \& $V$, and \re$\simeq$0.27$\pm$0.05\
in\ $I$ (or 1.8-2.5 kpc), pointing at best to a small color gradient. Its
position angle ($PA\sim$110\deg) is uncertain, but consistent with the
orientation of the aligned clouds ($PA\sim$ 95\deg), which appear to be
separated by a redder feature (\S 3.4 \& Fig. 2f). The $I$-band profile is
dominated by the continuum --- and least affected by the blue clouds which
contribute $\sim$8\% of the total $I$-band flux --- and is a somewhat better
$r^{1/4}$-fit, consistent with a crossing time of collapsing galaxies that
increases with radius (Fig. 1g in van Albada 1982). 

53W002's average color is $(B-I)\simeq$1.3. Its $(V-I)$ color of $\sim$0.7 is
less contaminated by the blue cloud. Within the errors, both colors are
relatively constant with radius, so that any color gradient must be small for
$r$\cle 1\arcspt 0 (\cle 0.3 mag across the $PC$ image). W91 \& W94b present
12-band (\Lya$UBVRIgriJHK$) photometry for 53W002 and surrounding objects, and
W94b, M97, \& Sc97 discuss spectral evolution model fits to these, and similar,
color-redshift data to constrain stellar population ages (defined as the onset
of the major visible starburst). The general blue colors in Fig. 2f suggests
that, with the exception of the region possibly affected by a ''dust lane'',
53W002 is not enormously reddened by dust (\S 3.4). Following these models, the
colors of the symmetric component of 53W002 --- if interpreted as coming from
stars only --- would suggest a stellar population with an average age of
$\sim$0.4 Gyr, which is of the same order as its dynamical time scale (\S 4).
Without the detection of a significant color gradient in the stellar population,
the current data cannot distinguish whether 53W002 formed through a sudden,
global halo collapse (\cf ELS62) or through rapid merging of many sub-galactic
sized units (\eg SZ78, P96). $HST/NICMOS$ $J,H$ images that bracket the 4000\AA\
break will help decide between these scenarios. Below we show that there may be
other processes triggering the (star)formation of 53W002.

\subsection {The nature of the blue clouds } 

The residuals after removing the symmetric pieces described in \S 3.1--3.3
trace the aligned blue ''clouds'' nicely. The larger cloud to the west --- in
the direction of the extended \Lya cloud (Fig. 2a \& W91) --- is quite extended
and vaguely triangular (Fig. 2e and insert in Fig. 1). It peaks 0\arcspt 45 west
of the nucleus and extends \cge 1\arcs from the core with an opening angle of
about 45$^\circ$ (Fig. 2b). In \Lya, it is dominated by a brighter ''arc'' about
0\arcspt 6 from the core (Fig. 2a). On the opposite side is a very small blue
object --- possibly a ''counter-cloud'' --- elongated perpendicular to the
nucleus-cloud direction and confined within 0\arcspt 2 from the core (Fig. 2e)
--- at the very limit of the $HST/PC$ resolution. These blue ''clouds'' comprise
13\% and 5\%, respectively, of the total F450W flux, and probably account for a
good fraction of both the size [\re(I)$\sim$1\arcspt 1] and elongation of 53W002
inferred from the pre-refurbished WFC images in $V$ \& $I$ (W92). Hence, we now
interpret 53W002's stellar light distribution as much smaller and less
elongated. Similar analysis of the PC-images in \Vband \& \Iband (with
correspondingly reduced angular resolution) shows the same basic components
(Fig. 2c \& 2d). The core and small extended region immediately surrounding it
are significantly bluer than their surroundings, as is the western extended
cloud (Fig. 2f). With the exception of the ''arc'' at the edge of the larger
cloud, the aligned components and the core are essentially free of \Lya
line-contamination (Figs. 2a--2f). The nucleus is a weak \Lya source,
contributing only about 20\% of the total \Lya flux (Fig. 2a). \Lya contributes
about 17\% of the overall $B$ light in a 2\arcspt 5 aperture, somewhat smaller
than the $\sim$30\% expected from the integrated spectrum (W91 \& P96). That is,
the \Lya line emission is no more concentrated than the continuum light, and is
thus not dominated by the nucleus. In the eastern cloud, \Lya contributes no
more than 10\% of the $B$-band flux, but the ''arc'' at the outer edge of the
western cloud contributes as much as 93\%. This is the only feature seen in the
\Lya image with significant contrast against the rest of the galaxy in terms of
equivalent width. These two blue clouds could represent: 

\si (1) Reflection of the AGN-light shining through a cone, including \Lya and
C-IV emission lines from gas lit up by the cone. The asymmetry in size and flux
between eastern and western clouds (Fig. 2b \& 2e) may represent obscuration or
genuine physical differences. The fact that we can see a much larger and
somewhat symmetric ground-based \Lya cloud (compare our Fig. 2a to Fig. 3 of
W91) argues against obscuration, although this extended \Lya gas could be mostly
in front of --- or away from --- any dust, and in part be unrelated to the AGN.
We note that two other z$\simeq$2.40 objects (Obj. 18 \& 19 of P96) are also AGN
with continuum reflection cones (see inserts in Plate 1), but with a relatively
stronger AGN component compared to the surrounding material (at the 50--80\%
level of the total flux). The presence of these reflection cones implies the
existence of a substantial amount of gas and/or dust well beyond the optical
extent of these galaxies (see \S 4). 

\si (2) A star-bursting region induced by 53W002's radio jet. Compared to the
spectral evolution models described by W94b, the much bluer colors of the cloud
--- if caused by stars --- would suggest a star-bursting region \cle $10^{8}$
years old. This is similar to the typical radio source lifetime, but younger
than the galaxy's dynamical time scale. 

A color map --- produced by matching the registration, sampling and resolution
of the \Bband \& \Iband images --- shows the color contrast between the inner
and outer regions of 53W002 (Fig. 2f). This map also provides an interesting
clue as to the possible origin of the smaller ``cloud'' seen near the nucleus in
the \Bband images: a red, almost linear feature appears to separate this smaller
cloud from the nucleus. In nearby galaxies, this would suggest an organized dust
lane. The color of the outskirts of this smaller cloud is $(B-I)\simeq$0.6 (at
r=0\arcspt 13), and for the red ``dust'' lane it is $(B-I)\simeq$1.0. Flux that
doesn't show up in these $PC$ components must be redder yet to match the total
ground-based (W91) and the global $HST$ color of $(B-I)\simeq$1.3 (which is
close to colors of the symmetric halo). If this feature is indeed a dust lane,
it would have a differential optical depth between 1300 and 2400 \AA\ ranging
from $\tau=0.75-1.5$ averaged over the resolution limit. This is rather mild by
standards of present-day galaxies. The visual or blue extinction expected for
this amount of far-UV extinction would be easy to miss in nearby radio galaxies,
so the total amount of dust required is not excessive for objects like 53W002,
which might be chemically younger and correspondingly more metal-poor.

\section {Discussion and conclusions } 


\subsection {Nature of the alignment effect in 53W002} 

The 8.4 GHz contours in Fig. 2 show that the radio source is {\it not}
smaller than the distance between the blue continuum clouds, and that the radio
jet coincides with the bright \Lya ''arc'' in the western cloud, suggesting that
jet-induced star-formation could indeed cause both blue clouds, except the outer
parts of the western cloud. The latter has a more ''jagged, triangular'' shape,
as the color image of Plate 1 shows, and so may be caused by AGN light in
reflection. A recent interferometric OVRO image of 53W002 in redshifted CO (J\ \
3--2) --- which has 3\arcs FWHM --- provides an important clue to the alignment
effect (S97). The CO was detected up to 2--3\arcs away on both sides of 53W002's
AGN, {\it and in the same direction as both blue $HST$ clouds and the extended
8.4 GHz radio source, but not perpendicular to this direction,} so that at least
the CO that coincides with the currently visible jet was likely deposited there
by physical processes related to this jet. Since Carbon and Oxygen had to be
formed in massive stars, jet-induced star-formation thus likely played a role in
53W002. The CO extends further in both directions than the two aligned blue
clouds and the current radio jet (Fig. 3 of S97 and Fig. 2a here). This is
harder to understand through star-formation from the current compact jet, which
is likely confined through a dense ISM in the galaxy (\cf de Vries \etal 1997),
unless the jet in the past managed to get further out of the galaxy. This might
be possible through holes in its dense ISM, if e.g. 53W002 formed through the
rapid merging of surrounding sub-galactic sized objects (P96). 53W002's
$r^{1/4}$-like stellar population is extended in the same direction as the radio
source, so that the jet possibly triggered a non-negligible fraction of 53W002's
mass to form stars in these two directions. As long as this all happened within
a few$\times 10^{8}$ years, there could have been just enough time for the
stellar population to settle into a possible $r^{1/4}$-like profile. Together
with the continuum and \Lya morphology of the blue clouds (\S 3.4 \& Fig. 2), it
thus appears that both reflection cones from an AGN (Fig. 1) and jet-induced
star-formation are responsible for the alignment effect in 53W002. 

We can compare the structure of 53W002 to the $HST$ images of powerful radio
galaxies of Longair, Best, \& R\" ottgering (1995) and Best, Longair, \& R\"
ottgering (1996, 1997). From a set of eight 3CR radio galaxies at $z \approx 1$,
they suggest that the morphologies change systematically with (projected) radio
source size, and interpret this as evidence for jet-induced star formation in
the aligned component. The radio galaxy 53W002 shares both properties with these
powerful 3CR sources: a compact component --- the AGN in the center of an
extended starlight distribution --- and a pair of clouds aligned with the
projected radio axis. Contrary to the 3CR sources, we can already see a fairly
relaxed symmetric distribution of starlight centered around 53W002's nucleus. In
hindsight, it is perhaps surprising to find the aligned component to be
important even at these low radio powers, but the continuity of structure with
the powerful radio galaxies is striking, and required the extra resolving power
of the $WFPC2/PC$ in the $B$-band to observe in detail. 


\subsection {53W002's gas+dust content, surrounding cluster, and possible
evolution} 

The measured \Lya fluxes and broad-band UBVRIgriJHK colors of 53W002
constrain its dust absorption (\AV\cle 0.2 mag) and star formation rate (SFR;
W91, W94b). Similar arguments have been made by P96 for the other 16 surrounding
blue $z=2.40$ candidates. The SFR of 53W002 is of order $\sim$100\Mo/year (W91),
and $\sim$5--10$\times$ less for the other $z=2.40$ candidates (P96). The total
stellar mass of 53W002 --- integrated over its assumed exponentially declining
SFR --- is $\sim 1.8\times 10^{11}$\Mo (W91). The $r^{1/4}$-like light-profile
of 53W002 suggests that at $z=2.39$ the object had already converted a
non-negligible fraction of its gas mass into stars --- rather efficiently on a
$\sim$0.4 Gyr time-scale --- suggesting a young early-type galaxy. The OVRO
CO-flux of 1.5 Jy/km/s implies $\sim 2.1\times 10^{11}$\Mo in gas for 53W002
alone (S97). The velocity widths of the CO clouds are $\sim$250 km/s (HWHM),
extending $\sim$1\arcspt 5 or $\sim$13 kpc on each side of 53W002 in the
direction of the blue clouds {\it and} of the extended radio source, and
possibly indicating a forming rotation curve (S97). These numbers imply an
enclosed Keplerian mass of 1.5--3.8$\times 10^{11}$\Mo, consistent with its
total stellar mass above (W91). Taken together, this means that 53W002's
$H_{2}$(+CO) gas-mass could be $\sim$30--60\% of its total (luminous+gas+dark)
mass. 

What could this mean for the evolution of 53W002? If all this gas settled into
disk stars within a few free-fall times ($\sim$1 Gyr), 53W002 could evolve into
an mid-type \Lstar spiral galaxy today (with $B/D$-ratio $\sim$0.5), or into an
earlier-type galaxy ($B/D$\cge 1) if it did not, and if a substantial fraction
remained neutral (as seen in some nearby ellipticals and merger remnants, \cf
Hibbard 1995). The small velocity dispersion (\cle 300 km/s) in P96's group of
$z=2.4$ objects with measured redshifts, and the small area (\cle 1 Mpc$^2$)
over which the 16 surrounding sub-galactic objects are seen, suggests that many
of these objects will likely merge into a few larger galaxies during the next
half Hubble time after $z\simeq 2.4$. Hence, while 53W002 may have formed with a
$r^{1/4}$-like profile during a relatively quick and sudden collapse that
started at $z\simeq$3 ($\sim$0.4 Gyr before $z=2.4$) --- possibly through
star-formation along its radio jet --- it also appears to be developing a
massive disk at $z\simeq2.39$ (S97). This disk may have completely settled
$\sim$1--2 Gyrs later (or at $z\sim$1.5), but possibly be destroyed again during
future mergers (at $z$\cle 1.5) with the surrounding sub-galactic sized objects
(P96), so that the radio galaxy 53W002 may end up as a massive early-type galaxy
today. 

\acknowledgments

We thank Doug VanOrsow, Ray Lucas and the STScI staff for their assistance,
Claudia Burg and Simon Driver for help in the image processing, Ed Fomalont for
his recent VLA positions, Jeff Hester, Dave Burstein and the referee for useful
suggestions. This work was supported by NASA grants GO-5308.0*-93A \&
GO-5985.0*-94A (to both RAW and WCK) from STScI, which is operated by AURA,
Inc., under NASA contract NAS5-26555. 


\ve \bn \cl{\bf Figure Captions} 

\figcaption[]{\label{fig1} (COLOR PLATE 1). Color image of the $HST/PC$
-exposures of the 53W002 field (12 hr in \Bband \& 5.7 hr in both \Vband \&
\Iband). The \Vband-band stack was rotated by 6.721\deg to match \Bband \&
\Iband (\& \Lyaband), resulting in slanted borders, but small enough for all
mosaics to cover roughly the same region (see D95a). North is 140.3\deg
counterclockwise from vertical. This $PC$ image covers $32\times 32$\arcs, and
has $\sim$0\arcspt 06 resolution (FWHM). The 3-$\sigma$ point source sensitivity
is 28.8 mag in \Bband, 29.5 in \Vband, 28.3 in \Iband, and 26.6 mag in the 11.25
hr \Lyaband images (Fig. 2a). The 1-$\sigma$ SB-sensitivity (per pixel) is
26.7\magarc in \Bband, 26.8 in \Vband, 26.2 in \Iband, and 25.2 in \Lyaband. The
bright red object ''S'' is the only known star in the image. Other objects are
labelled as in W91, W92, W94b, \& P96. The inner light-profile of 53W002 is
fairly regular, with a brighter blue ''cloud'' prominent to the upper-left
($\simeq$west; see Fig. 2). The WF2 inserts of the two z=2.40 AGN (No. 18 \& 19;
P96) are shown at the same scale as the 53W002 insert, and show similar
continuum cones. }

\figcaption[]{\label{fig2} (GLOSSY PLATE 2). Grey scale images of $66\times 66$
pixels ($\simeq$3\arcspt 0$\times$3\arcspt 0) of the PC-exposures on 53W002: (a)
15$\times$2700s orbits in \Lyaband; (b) 24$\times$2400s in \Bband; (c)
12$\times$1700s in \Vband; (d) 12$\times$1700s in \Iband; (e) equal to (b) after
subtraction of a central point source (with 25\% of the total $B$-band flux) and
of the best fit $r^{1/4}$-like light-profile (Fig. 3 \& \S 3.3); (f) the $(B-I)$
color image as the ratio of (b)/(d). The VLA 8.4 GHz contours of W91 and the CO
peaks of S97 are superimposed in (a) \& (d) (see text). The AGN is located at
each panel's central pixel. {\it Note:} the low SB and ''arclike'' feature of
the \Lya cloud in (a); the $r^{1/4}$-like bulge and brighter blue cloud in
(b)--(d); the brighter {\it and} fainter blue clouds in (e) where the symmetric
parts of 53W002 have been subtracted; and the blue (dark) central AGN and the
small reddish (white) linear feature separating both blue clouds in (f). The
radio source covers most of the blue clouds plus the \Lya ''arc'', suggesting
jet-induced starformation; the remainder of the continuum and \Lya cloud may be
reflected AGN light. }

\figcaption[]{\label{fig3} SB profiles in \Bband, \Vband \& \Iband vs. radius
(in $r^{1/4}$ units) for the extended component of the radio galaxy 53W002 at
$z=2.390$ in the quadrant that is relatively free of the blue clouds. The \Vband
profile was moved up by --0\magpt 2 for clarity. A central AGN component was
subtracted in each filter, using star ''S'' as PSF (Fig. 1). Short dashes
represent the 25\% AGN component in $B$, which affects the profile to the left
of the vertical dashed line ($r^{1/4}$\cle 0.5; $r$\cle 0\arcspt 06). Formal
errors reflect photon statistics only. Horizontal dotted lines indicate
2$\sigma$ sky-subtraction errors, which affect the profiles for $r^{1/4}$\cge
0.75--0.8 ($r$\cge 0\arcspt 4). For 0.5\cle $r^{1/4}$\cle 0.75 (0\arcspt 06\cle
$r$\cle 0\arcspt 4), 53W002's profile is consistent with an $r^{1/4}$-law,
especially if the AGN contribution is $\le$20\%, but at large radii it is
affected by the blue clouds. }

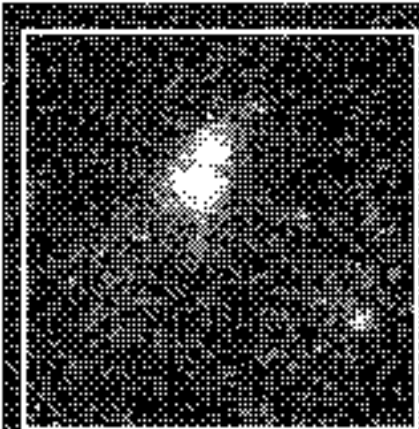
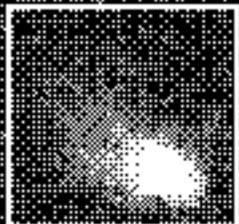
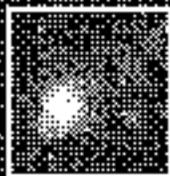
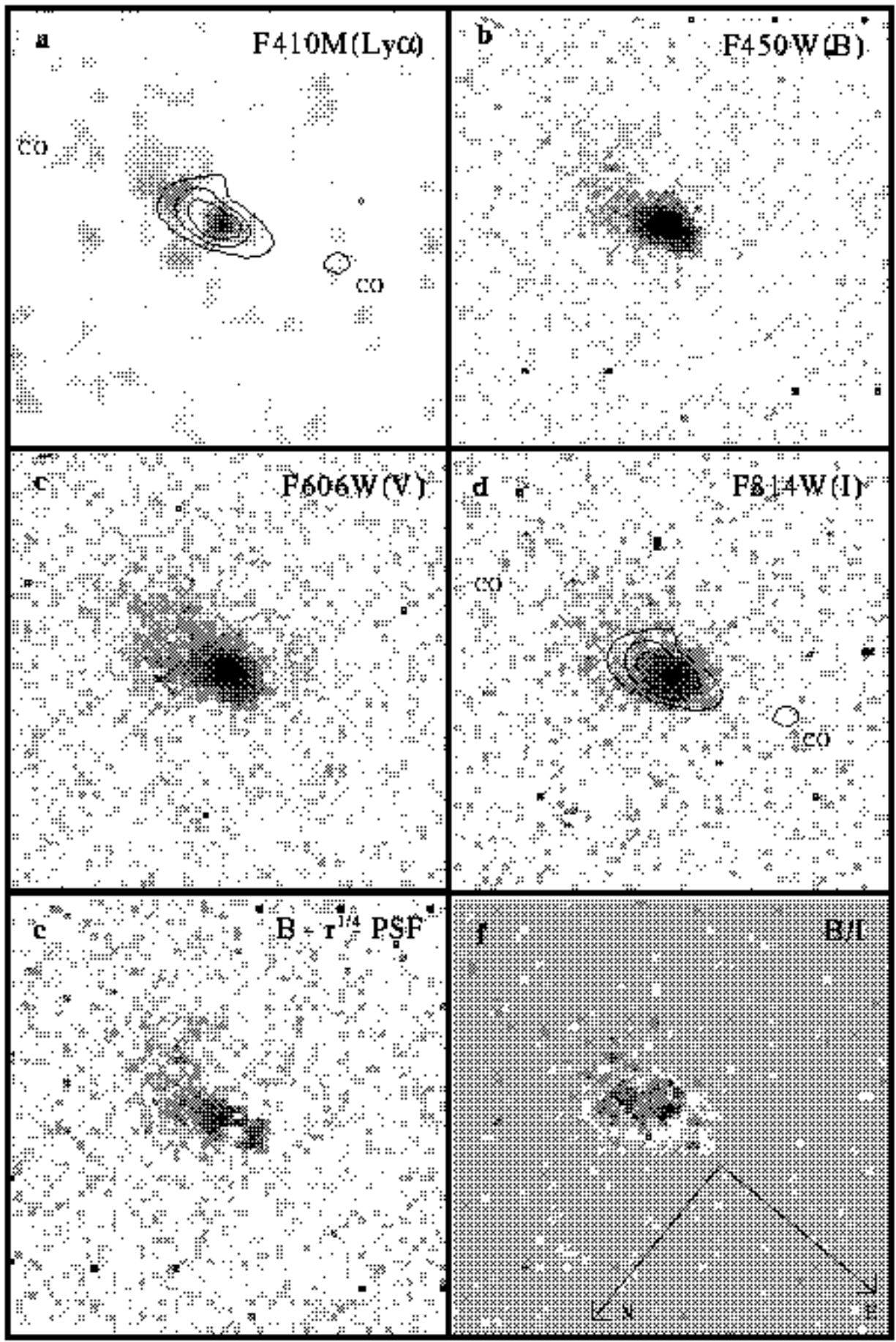
\end{document}